\documentclass[aps,prd,preprintnumbers,nofootinbib,onecolumn]{revtex4}

\usepackage{bm}
\usepackage{latexsym}
\usepackage{dcolumn}
\usepackage{amsmath,amsfonts,multirow,amssymb}
\usepackage{graphicx,epsfig}
\usepackage{psfrag}
\usepackage{amsthm}

\def\be {\begin{equation}}
\def\ee {\end{equation}}
\def\bea {\begin{eqnarray}}
\def\eea {\end{eqnarray}}
\def\bc {\begin{center}}
\def\ec {\end{center}}
\def\bfg {\begin{figure}}
\def\efg {\end{figure}}
\def\bi {\begin{itemize}}
\def\ei {\end{itemize}}
\def\nn {\nonumber}

\def\la {\label}
\def\le {\left}
\def\ri {\right}

\def\bes {\begin{split}}
\def\ees {\end{split}}

\def\beq{\begin{equation}}
\def\eeq{\end{equation}}
\def\br{\begin{eqnarray}}
\def\er{\end{eqnarray}}
\newcommand{\eel}[1] {\label{#1}\end{equation}}
\theoremstyle{plain}
\newtheorem{thm}{Theorem}

\newtheorem{defn}{Definition}
\newtheorem{eg}{Example}
\theoremstyle{Remark}

\newcommand{\bdm}{\begin{displaymath}}
\newcommand{\edm}{\end{displaymath}}

\begin{document}

\title{Generalized Uncertainty Principle and Self-Adjoint Operators}

\author{Venkat Balasubramanian $^{1}$} \email[email: ]{vbalasu8@uwo.ca}
\author{Saurya Das $^{2}$} \email[email: ]{saurya.das@uleth.ca}
\author{Elias C. Vagenas \vspace{0.5cm}$^{3}$} \email[email: ]{elias.vagenas@ku.edu.kw}

$\vspace{2cm}\\$

\affiliation{$^1$~Department of Applied Mathematics, University of
Western Ontario London, Ontario N6A 5B7, Canada\\}

\affiliation{$^2$~Theoretical Physics Group, Dept. of Physics and
Astronomy, University of Lethbridge, 4401 University Drive,
Lethbridge, Alberta T1K 3M4, Canada \\}

\affiliation{$^3$~Theoretical Physics Group, Department of Physics, Kuwait University, 
P.O. Box 5969, Safat 13060, Kuwait}

\begin{abstract}
\vspace{3ex}
\par\noindent
In this work we explore the self-adjointness of the GUP-modified momentum
and Hamiltonian operators over different domains. In particular, we utilize the theorem by
von-Newmann for symmetric operators in order to determine
whether the momentum and Hamiltonian operators are  self-adjoint or not, 
or  they have self-adjoint extensions over the given domain. In addition, a simple example of the Hamiltonian 
operator describing a particle in a box is given. The  solutions of the boundary conditions that describe the 
self-adjoint extensions of the specific Hamiltonian operator are obtained.

\end{abstract}

\maketitle

\newpage
\section{Introduction}

\par\noindent
One of the old, vexing, and still unsolved problems of Theoretical Physics is that of merging gravity with quantum field theory. 
Nowadays, the incarnation of this combination is the several theories of quantum gravity (such as String theory) that we have. One of the new features predicted by these theories  is the minimal measurable length (for a recent review see \cite{AmelinoCamelia:2008qg}, and references there in). This feature led to the 
generalization of the Heisenberg Uncertainty Principle, i.e. the Generalized Uncertainty Principle (henceforth abbreviated to GUP). In addition, GUP can also be considered as an outcome of modifications/corrections  to the conventional Heisenberg algebra satisfied by the two canonically conjugate observables:
position $\textbf{x}$ and momentum $\textbf{p}$. According to String Theory, the conventional 
Heisenberg algebra gains an extra term which is quadratic in momentum $\textbf{p}$, in the Planck regime.
In \cite{Kempf:1994su}, the authors used the following modified
Heisenberg algebra consistent with String Theory
\be
\label{eqn1}
\left[\hat{x}, \hat{p}\right] = i\hbar\left(1+\beta p^{2}\right)~.
\ee
\par\noindent
Black Hole physics, and Doubly Special Relativity (DSR) propose a correction in the Planck regime with the extra term
 to be linear in momentum $\textbf{p}$. 
In \cite{Ali:2009zq,Ali:2010yn}, the authors considered modifications to the conventional Heisenberg algebra,
which includes both linear and quadratic terms in momentum, namely
\be
\label{eqn2}
{[\textbf{$\hat{x}$},\textbf{$\hat{p}$}]} = i\hbar \left[1 - 2 \alpha p + 4 \alpha^{2} p^{2} \right]~.
\ee
\par\noindent
Utilizing the following uncertainty relationship satisfied by any two operators $\hat{A}$ and $\hat{B}$
\be
\label{eqn3}
\Delta \hat{A} \Delta \hat{B} \geq \frac{1}{2}\left\|\left<\left[\hat{A}, \hat{B}\right]\right>\right\|
\ee
\par\noindent
where $\Delta \hat{A}$ and $\Delta \hat{B}$ stand for the  standard deviations of the corresponding operators, 
a GUP between position $x$ and momentum $p$ is obtained
\be
\label{eqn4}
\Delta x\Delta p \geq \frac{\hbar}{2}\left[1-2\alpha\left<p\right>+4\alpha^{2}\left<p^{2}\right>\right]
\ee
\par\noindent
where  $l_{pl} \approx 10^{-35}$ m is the Planck length,  $\alpha = \alpha_{0}~l_{pl}/ \hbar $ is the 
GUP parameter, and it is normally assumed that $\alpha_{0} $ is of order $1$. In \cite{Ali:2009zq,Ali:2010yn}, the authors suggested an upper bound on $\alpha_{0}$ by stating that its value cannot exceed $10^{17}$ which is  precisely the electroweak length scale. This prediction comes
about due to the fact that if $\alpha_{0}$ were to be any larger, such an intermediate length scale would have been observed.
Now, it is easy to verify the fact that the above two equations, i.e., Eqs.(\ref{eqn2}) and (\ref{eqn4}),  
predict a minimum uncertainty in position, i.e,  $\Delta x_{min}$, and a maximum uncertainty in momentum, i.e., $\Delta p_{max}$,\footnote{It is noteworthy that DSR theories also introduce the features of minimal measurable length and  maximum measurable momentum \cite{Magueijo:2001cr,AmelinoCamelia:2010pd}.}
\bea
\Delta x_{min} &\propto & \alpha_{0}l_{pl} \\
\Delta p_{max} &\propto &\frac{M_{pl}c}{\alpha_{0}}
\label{eqn5}
\eea
\par\noindent
where $M_{pl}$ is the Planck Mass. It can be shown that the following representations 
of the position and momentum operators satisfy the modified Heisenberg  algebra given by Eqn.(\ref{eqn2})
\bea
\hat{x} &=& \hat{x}_{0} \\
\hat{p} &=& \hat{p}_{0}\left(1-\alpha\hat{p}_{0}+2\alpha^{2}\hat{p}^{2}_{0}\right)
\label{eqn6}
\eea
\par\noindent
with $\hat{x}_{0}$ and $\hat{p}_{0}$ satisfying the ordinary canonical commutation relations
$\left[\hat{x}_{0}, \hat{p}_{0}\right] = i\hbar$. Here, $\hat{p}_{0}$ can be interpreted as being
the total momentum of a particle at low energies and having the standard representation, namely,  in one dimension, 
\be
\hat{p}_{0} = -i\hbar\frac{d}{dx}~.
\label{eqn7}
\ee
\par\noindent
Now, considering any non-relativistic Hamiltonian of the form
\be
\hat{H} = \frac{\hat{p}^{2}}{2m} + V(\textbf{r})
\label{eqn8}
\ee
\par\noindent
we see that, due to Eq.(\ref{eqn6}), i.e., the GUP-modified momentum operator, every Hamiltonian of the form of Eq.(\ref{eqn8})
obtains higher order terms in $\alpha$ and $\hat{p}_{0}$. A simple substitution of the new momentum operator into
Eq.(\ref{eqn8}) yields 
\be
\hat{H} = \frac{\hat{p}_{0}^{2}}{2m}-\frac{\alpha}{m}\hat{p}_{0}^{3}+\frac{5\alpha^{2}}{2m}\hat{p}_{0}^{4}-\frac{2\alpha^{3}}{m}\hat{p}_{0}^{5}+
\frac{2\alpha^{4}}{m}\hat{p}_{0}^{6} + V(\textbf{r})~.
\ee
\par\noindent
A last but not least comment for the scope of this paper is in order here. It is well-known that in Physics we are interested in observables with real eigenvalues, which are guaranteed only for self-adjoint operators, barring potentially pathological cases that are of no apparent interest to our purposes. Thus, it is very important to know if an operator is self-adjoint or not, or if it has self-adjoint extensions over specific domains. 
This is the reason that in this paper we investigate the self-adjointness of the GUP-modified momentum and Hamiltonian operators, characterized by the powers of $\alpha$ up to $\mathcal{O}(\alpha^{2})$.
The remainder of the paper is structured as follows. 
In section II, we briefly present some mathematical tools which will be used in next sections. In section III, 
we show in which domains the GUP-modified  momentum operator is self-adjoint, in which domains it is not self-adjoint, and in which domains it has infinitely many self-adjoint extensions. We follow this analysis when the GUP-modified momentum operator has a linear term in the GUP parameter $\alpha$, when it has a quadratic term in $\alpha$, and when both terms, i.e., the linear and the quadratic in $\alpha$, are present. In section IV, we perform the same analysis as presented in section III for the case of the Hamiltonian operator. In section VI, we present a simple example of GUP-modified Hamiltonian with a linear term in $\alpha$. Finally, in section VI, we briefly summarize  our results.

\section{Mathematical Preliminaries}
\par\noindent
In this section, we briefly present  all necessary definitions, theorems, and lemmas concerning linear operators 
\cite{IntroHilbertSpacesApps, introFuncAnalysis, funcAnalysis1, mthdsMathPhys}.

\begin{defn}
Let $\mathcal{V}$ a normed vector space. A linear mapping $\hat{A}:\mathcal{V}\rightarrow\mathcal{V}$ is called a \textbf{linear operator} in
$\mathcal{H}$. The subspace of elements $x \in \mathcal{H}$ for which $\hat{A}x$ is defined is termed as the \textbf{domain} of $\hat{A}$ and
is denoted as $\mathcal{D}_{\hat{A}}$. The \textbf{range} of $\hat{A}$ is the set of all elements $y \in \mathcal{V}$ such that $y = \hat{A}x$ holds,          and is denoted as $\mathcal{R}_{\hat{A}}$.
\end{defn}


\par\noindent
At this point, it should be stressed that from now onwards we use the words {\bf linear operator} and {\bf operator} interchangeably.

\begin{defn}
Let $\mathcal{V}$ be a normed vector space. Any linear mapping of $\mathcal{V}$ into $\mathcal{V}$ itself is called a \textbf{bounded} operator
if the $\|\hat{A}\| < \infty$. The norm of an operator is defined as follows:
\begin{equation}
\|\hat{A}\| = \sup_{x \in \mathcal{D}_{\hat{A}}\\\|x\| \neq 0} \frac{\|\hat{A}x\|}{\|x\|}~.
\end{equation}
\end{defn}

\par\noindent
The above simply implies that we can always find a positive real constant, say $\mathcal{M}$, such that $\|\hat{A}x\| \leq \mathcal{M}\|x\|$.
In the case when no such constant exists, we term the operator to be \textbf{unbounded}. \\

\par\noindent
In this section, we focus more on the properties of unbounded operators since most of the operators encountered in 
Physics such as the momentum and the Hamiltonian are unbounded operators.

\begin{eg}
Consider the differential operator $\frac{d}{dx}$ to be defined on the space of all differentiable functions on some interval
$\left[a,b\right] \subset \mathbb{R}$, which is a subspace of $\mathcal{L}^{2}(\left[a,b\right])$. Suppose we consider a sequence of functions
$f_{n}(x) = \sin(nx)$, n = 1,2,3,\ldots, defined on $\left[-\pi,\pi\right]$. Then
\be
\|f_{n}\| = \sqrt{\int_{-\pi}^{\pi} \left(\sin nx\right)^{2} dx} = \sqrt{\pi} < \infty
\ee
\par\noindent
and
\be
\|\frac{d}{dx}f_{n}\| = \sqrt{\int_{-\pi}^{\pi} \left(n \cos nx\right)^{2} dx} = n \sqrt{\pi}~.
\ee
\par\noindent
From the above we see that, there is no real constant that can set an upper bound on $\|\frac{d}{dx}f_{n}\|$,
hence we see that the differential operator is unbounded.
\end{eg}

\par\noindent
Since the GUP-modified momentum and Hamiltonian operators are differential operators, we will completely work
with unbounded operators and for this reason the following definitions and theorems concern only unbounded operators.

\begin{defn}
Let $\mathcal{H}$ be a normed vector space. Let $\hat{A}$ be an operator such that $\mathcal{D}_{\hat{A}} \in \mathcal{H}$.
Then $\hat{A}$ is said to be densely defined if $\mathcal{D}_{\hat{A}}$ is dense in $\mathcal{H}$, i.e $\forall \psi \in \mathcal{H}$,
$\exists \in \mathcal{D}_{\hat{A}}$ a  sequence $\phi_{n}$ which in norm converges to $\psi$
\end{defn}

\begin{defn}
An operator $\hat{A}: \mathcal{H} \rightarrow \mathcal{H}$, with domain $\mathcal{D}_{\hat{A}} \in \mathcal{H}$, is said to be \textbf{closed}
if its graph $\Gamma{(\hat{A})}$
\be
\Gamma{(\hat{A})} = \left\{\left(x,y\right) | x \in \mathcal{D}_{\hat{A}}, y = \hat{A}x\right\}
\ee
\par\noindent
is closed in the normed space $\mathcal{H}\, \mathrm{x}\, \mathcal{H}$.\\
\end{defn}


\par\noindent
For an unbounded operator, one can define the corresponding \textit{adjoint} operator in the following way:

\begin{defn}
The adjoint, $\hat{A}^{\dagger}$ of an unbounded operator $\hat{A}$ defined in a Hilbert space, is defined as
\be
\forall x \in \mathcal{D}_{\hat{A}}, \forall y \in \mathcal{D}_{\hat{A}^{\dagger}} \, \, \,\, \,
\left<y|\hat{A}x\right> = \left<\hat{A}^{\dagger}y|x\right>~.
\ee
\end{defn}

\par\noindent
Since,  we only deal with dense domains, we  will consider only densely defined unbounded operators.
\begin{thm}
\label{ThmProof1}
If $\hat{A}$ is a densely defined operator, then its adjoint $\hat{A}^{\dagger}$ is closed
\end{thm}
\begin{defn}
\par\noindent
\begin{enumerate}

\item Let $\hat{A}$ be an operator defined in the Hilbert space, $\mathcal{H}$. The $\hat{A}$ is called \textbf{Hermitian} or
\textbf{symmetric} if, $\forall x, y \in \mathcal{D}_{\hat{A}}$,
\be
\left<y | \hat{A}x\right> = \left<\hat{A}y | x\right>~.
\ee
\item An operator $\hat{A}$ defined in a Hilbert space $\mathcal{H}$ is said to be \textbf{self-adjoint}
if it is densely defined over its domain and in form, $\hat{A} = \hat{A}^{\dagger}$.

\end{enumerate}

\end{defn}

\par\noindent
{\bf Note. }
The equality $\hat{A} = \hat{A}^{\dagger}$, apart from the equality in form of the two operators,
also implies that the respective domains of the operators should also coincide, i.e., $\mathcal{D}_{\hat{A}} = \mathcal{D}_{\hat{A}^{\dagger}}$.\\

\par\noindent
In the case of bounded operators, one need not concern with the equality of domains because, the domain of a
densely defined bounded operator can always be extended to the entire vector space. Therefore, a bounded Hermitian operator is also
self-ajdoint. However in the unbounded case, the situation is a little bit more subtle, since the operator being symmetric
doesn't imply self-adjointness. We now describe the \textit{von Neumann's Theorem} which is an indispensable tool in the
analysis of self-adjointness of operators.\\

\par\noindent
{\bf Note. } From now onwards we consider only  unbounded operators.\\

\begin{defn}
Let $\hat{A}$ be a symmetric operator. Let
\bea
\mathcal{K}_{+} &=& \ker(i-\hat{A}^{\dagger}) \\
\mathcal{K}_{-} &=& \ker(i+\hat{A}^{\dagger})
\eea
\par\noindent
where $\mathcal{K}_{+}$ and  $\mathcal{K}_{-}$ are called \textbf{deficiency subspaces} of $\hat{A}$ and their dimensions,
i.e., $n_{+} = \dim\left[\mathcal{K}_{+}\right]$  and $n_{-} = \dim\left[\mathcal{K}_{-}\right]$ are referred to as the \textbf{deficiency indices} of $\hat{A}$.
\end{defn}

\par\noindent
{\bf Note. }The deficiency indices of $\hat{A}$ can be any positive integer and even infinite.\\

\begin{defn}
Let $\hat{A}$ be an operator in a Hilbert space, $\mathcal{H}$. We say $\hat{B}$ is an \textbf{extension} of
$\hat{A}$ if the following conditions hold

\begin{itemize}

\item $\mathcal{D}(\hat{A}) \subset \mathcal{D}(\hat{B})$

\item $\hat{A} \phi = \hat{B}\phi, \forall \phi \in \mathcal{D}(\hat{A})$

\end{itemize}

\par\noindent
i.e. $\hat{A} \subset \hat{B}$.
\end{defn}

\par\noindent
Given an operator $\hat{A}$ in a Hilbert space and say $\hat{B}$ is a closed symmetric extension of the same,
then the following are true

\begin{itemize}

\item For $\phi \in \mathcal{D}(\hat{B}^\dagger)$
\be
(\psi, \hat{B}^\dagger\phi) = (\hat{B}\psi, \phi) = (\hat{A}\psi, \phi)
\ee
for all $\psi \in \mathcal{D}(\hat{A})$. Thus, from the above we see that $\phi \in \mathcal{D}(\hat{A^{\dagger}})$ and
$\hat{B}^\dagger \phi = \hat{A}^\dagger \phi$ so
\be
\hat{A} \subset \hat{B} \subset \hat{B}^\dagger \subset \hat{A}^\dagger .
\ee
\end{itemize}

\begin{center}
\textbf{von Neumann's Theorem}
\end{center}
\begin{thm}
Let $\hat{A}$ be a closed symmetric operator with deficiency indices $n_{+}$ and $n_{-}$. Then,
\begin{itemize}

\item $\hat{A}$ is self-adjoint if and only if $(n_{+}, n_{-}) = (0,0)$.

\item $\hat{A}$ has self-adjoint extensions if and only if $n_{+} = n_{-}$. These extensions are parametrized by an $n\,\mathrm{x}\,n$ unitary matrix.

\item If $n_{+} \neq n_{-}$, the $\hat{A}$ has no self-adjoint extensions.

\end{itemize}
\end{thm}
\section{GUP-modified Momentum}

\par\noindent
In this section we will apply von Neumann's theorem to the GUP-modified momentum operator  \cite{Bonneau:1999zq}. 
For this reason, we first have to determine the functions $\psi_{\pm}(x)$ which satisfy the equation
\be
\hat{p} \psi_{\pm}(x) = \pm i\frac{\hbar}{d}\psi_{\pm}(x)
\label{self-adjoint-eqn}
\ee
\par\noindent
where $d$ is a  positive constant   introduced for dimensional reasons and which is homogeneous to some length. \\
For the case of the total momentum of a particle at low energies and employing the standard representation, 
Eq.(\ref{self-adjoint-eqn}) becomes
\be
-i\hbar\frac{d\psi_{\pm}(x)}{dx} = \pm i\frac{\hbar}{d}\psi_{\pm}(x)~.
\ee
It is easily seen that a solution to the above equation reads
\be
\psi_{\pm}(x) = C_{\pm}\exp\left[\mp\frac{x}{d}\right]~.
\ee
\par\noindent
Over different domains, the deficiency indices and the self-adjointness of the momentum operator are described as follows:

\begin{itemize}

\item $\mathcal{D}(\hat{p}_{0}) = \mathcal{L}^{2}(\mathbb{R})$ : $(n_{+},n_{-})=(0,0)$ and thus  the operator is self-adjoint.

\item $\mathcal{D}(\hat{p}_{0}) = \mathcal{L}^{2}[0,\infty)$ : $(n_{+},n_{-})=(1,0)$ and thus  the operator is not self-adjoint.

\item $\mathcal{D}(\hat{p}_{0}) = \mathcal{L}^{2}([0,L])$ : $(n_{+},n_{-})=(1,1)$ and thus the operator has infinitely many self-adjoint
extensions parametrized by a $U(1)$ group.

\end{itemize}
\par\noindent
For future convenience, we will make all variables dimensionless. For this reason, using  the quantity $\alpha\hbar$ 
which is the physical length scale introduced in GUP, we define a new dimensionless parameter $\rho$ as follows 
\bea
\rho = \frac{x}{\alpha\hbar}~.
\eea
Therefore, the momentum operator $\hat{p}_{0}$ expressed in terms of the new variable $\rho$ now reads
\bea
\hat{p}_{0} = -i\hbar\frac{d}{dx} = -\frac{i}{\alpha}\frac{d}{d\rho}~.
\eea
%

\subsection{Momentum operator with linear term in $\alpha$}

\par\noindent
Let us now consider as momentum operator in Eq.(\ref{self-adjoint-eqn}), the GUP-modified 
momentum operator which is linear in $\alpha$, namely
\bea
\hat{p} &=& \hat{p}_{0}-\alpha\hat{p}_{0}^{2}.
\eea
\par\noindent
First we write  Eq.(\ref{self-adjoint-eqn}) for the function $\psi_{+}$ and we get
\be
\left(\hat{p}^{\dagger} - \frac{i}{\alpha}\right)\,\psi_{+}(x) = 0 \notag 
\ee
\par\noindent
which in terms of the dimensionless parameter $\rho$ takes the form
\be
\left(\frac{d^{2}}{d\rho^{2}}-i\,\frac{d}{d\rho}-i\right)\,\psi_{+}(\rho) = 0 \notag ~.
\ee

\par\noindent
The characteristic equation of the above differential equation is
\be
\lambda^{2}-i\lambda-i = 0
\ee
\par\noindent
and so its roots are of the form: $\lambda^{1}_{+} = + 0.6 + 1.3\,i $ and $ \lambda^{2}_{+} = - 0.6 - 0.3\,i~$. 
Therefore, we obtain two  linearly independent solutions for $\psi_{+}$ 
\bea
\psi_{+}^{1}(\rho) &\propto & \exp\left[(+ 0.6 + 1.3\,i)\rho\right] \\
\psi_{+}^{2}(\rho) &\propto & \exp\left[(- 0.6 - 0.3\,i)\rho\right]~.
\eea

\par\noindent
Furthermore, over the different domains the deficiency index $n_{+}$  of the GUP-modified 
momentum operator which is linear in $\alpha$  is given as follows:

\begin{itemize}

\item $\mathcal{D}(\hat{p}) = \mathcal{L}^{2}(-\infty,\infty)$


In this domain, none of the functions, i.e., $\psi_{+}^{1}(x)$ and $\psi_{+}^{2}(x)$, is square integrable. 
So these functions do not belong to this space and thus $n_{+} = 0$.\\


\item $\mathcal{D}(\hat{p}) = \mathcal{L}^{2}(0,\infty)$

In this domain, only $\psi_{+}^{2}(x)$ has finite norm. So this is the only solution from the set which is
square integrable and thus $n_{+} = 1$.\\

\item $\mathcal{D}(\hat{p}) = \mathcal{L}^{2}([a,b])$

Over the finite interval both solutions, i.e., $\psi_{+}^{1}(x)$ and $\psi_{+}^{2}(x)$, are square integrable. 
So these functions belong to this space and thus $n_{+} = 2$.

\end{itemize}

\par\noindent
Second we write  Eq.(\ref{self-adjoint-eqn}) for the function $\psi_{-}$ and we get
\be
\left(\hat{p}^{\dagger}+\frac{i}{\alpha}\right)\,\psi_{-}(x) = 0 \notag 
\ee
\par\noindent
which in terms of the dimensionless parameter $\rho$ takes the form
\be
\left(\frac{d^{2}}{d\rho^{2}}-i\,\frac{d}{d\rho}+i\right)\,\psi_{-}(\rho) = 0 \notag~.
\ee

\par\noindent
The characteristic equation of the above differential equation is
\be
\lambda^{2}-i\lambda+i = 0
\ee

\par\noindent
and so its roots are of the form: $\lambda^{1}_{-} = + 0.6 - 0.3\,i $ and $\lambda^{2}_{-} =  - 0.6 + 1.3\,i $ .
Therefore, we obtain two  linearly independent solutions for $\psi_{-}$ 
\bea
\psi_{-}^{1}(\rho) &\propto & \exp\left[(+0.6 - 0.3\,i)\rho\right] \\
\psi_{-}^{2}(\rho) &\propto &  \exp\left[(- 0.6+ 1.3\,i)\rho\right]~.
\eea
\par\noindent
Furthermore, over the different domains the deficiency index $n_{-}$  of the GUP-modified 
momentum operator which is linear in $\alpha$  is given as follows:

\begin{itemize}

\item $\mathcal{D}(\hat{p}) = \mathcal{L}^{2}(-\infty,\infty)$

In this domain, none of the functions, i.e., $\psi_{-}^{1}(x)$ and $\psi_{-}^{2}(x)$, is square integrable. 
So these functions do not belong to this space and thus $n_{-} = 0$.\\

\item $\mathcal{D}(\hat{p}) = \mathcal{L}^{2}(0,\infty)$

In this domain, only $\psi_{-}^{2}(x)$ has finite norm. So this is the only solution from the set which is
square integrable and thus $n_{-} = 1$.\\

\item $\mathcal{D}(\hat{p}) = \mathcal{L}^{2}([a,b])$

Over the finite interval both solutions, i.e., $\psi_{+}^{1}(x)$ and $\psi_{+}^{2}(x)$, are square integrable. 
So these functions belong to this space and thus $n_{+} = 2$.

\end{itemize}

\par\noindent
Finally, employing von Newmann's theorem,  the self-adjointness of the GUP-modified momentum operator which is linear in $\alpha$ 
is described as follows :

\begin{itemize}

\item $\mathcal{D}(\hat{p}) = \mathcal{L}^{2}(-\infty,\infty) : (n_{+},n_{-}) = (0,0)$ and thus the momentum operator is self-adjoint.

\item $\mathcal{D}(\hat{p}) = \mathcal{L}^{2}(0,\infty) : (n_{+},n_{-}) = (1,1)$ and thus the momentum operator has infinitely many self-adjoint extensions.

\item $\mathcal{D}(\hat{p}) = \mathcal{L}^{2}([a,b]): (n_{+},n_{-}) = (2,2)$  and thus the momentum operator has infinitely many self-adjoint extensions.

\end{itemize}

\par\noindent
It is noteworthy that in \cite{Kempf:1994su} the authors utilized a similar modified momentum operatorÊ with a linear term in $\alpha$ to compute the scalar product between two wave functions Êof the momentum operator. In this context, the domain of the momentum operator was Ê$(-\infty, \infty)$. In order the operator to remain symmetric and thus self-adjoint in this calculation, the authors introduced a new measure in the integration so they changed the functional analysis of the operator. In our analysis, the measure remains the same and we change the domain of the momentum operator thus depending of the domain our momentum operator is self-adjoint. To our knowledge there is not a generic mathematical way to relate the results of these two different approaches.Ê However, this could form the part of a separate investigation.
\subsection{Momentum Operator with quadratic term in $\alpha$}
\par\noindent
We now consider as momentum operator in Eq.(\ref{self-adjoint-eqn}), the GUP-modified 
momentum operator which is quadratic in $\alpha$, namely
\bea
\hat{p} &=& \hat{p}_{0}+2\alpha^{2}\hat{p}^{3} \notag 
\eea
and we adopt the analysis of the previous subsection in order to study the self-adjointness of this 
GUP-modified momentum operator.

\par\noindent
First we write  Eq.(\ref{self-adjoint-eqn}) for the function $\psi_{+}$ and we get
\be
 \left(\hat{p}^{\dagger}-\frac{i}{\alpha}\right) \psi_{+}(x) = 0 \notag 
\ee
\par\noindent
which in terms of the dimensionless parameter $\rho$ takes the form
\be
 \left(2\,\frac{d^{3}}{d\rho^{3}}-\frac{d}{d\rho}-1\right)\,\psi_{+}(\rho) = 0 \notag~.
\ee
\par\noindent
The characteristic equation of the above differential equation is
\be
2\,\mu^{3}-\mu-1 = 0
\ee
\par\noindent
and so its roots are of the form: $\mu^{1}_{+} = 1$, $\mu^{2}_{+}  = -0.5 - 0.5\,i$, and $\mu^{3}_{+}  = -0.5 +0.5\,i$~. 
Therefore, we obtain three  linearly independent solutions for $\psi_{+}$ 
\bea
\psi_{+}^{1}(\rho) &\propto & \exp\left(\rho\right) \\
\psi_{+}^{2}(\rho) &\propto &  \exp\left[\left(- 0.5 - 0.5 \,i\right)\rho\right]\\
\psi_{+}^{3}(\rho) &\propto & \exp\left[\left(-0.5 + 0.5 \,i\right)\rho\right]~.
\eea
\par\noindent
Furthermore, over the different domains the deficiency index $n_{+}$  of the GUP-modified 
momentum operator which is in $\alpha$  is given as follows:
\begin{itemize}

\item $\mathcal{D}(\hat{p}) = \mathcal{L}^{2}(-\infty,\infty)$

In this domain, none of the functions, i.e., $\psi_{+}^{1}(x)$, $\psi_{+}^{2}(x)$, and $\psi_{+}^{3}$, is square integrable. 
So these functions do not belong to this space and thus $n_{+} = 0$.\\

\item $\mathcal{D}(\hat{p}) = \mathcal{L}^{2}(0,\infty)$

In this domain, only $\psi_{+}^{2}(x)$ and $\psi_{+}^{3}$ have finite norm. So there are only two solutions 
from the set which are square integrable and thus $n_{+} = 2$.\\

\item $\mathcal{D}(\hat{p}) = \mathcal{L}^{2}([a,b])$

Over the finite interval both solutions, i.e., $\psi_{+}^{1}(x)$, $\psi_{+}^{2}(x)$, and $\psi_{+}^{3}$, are square integrable. 
So these functions belong to this space and thus $n_{+} = 3$.

\end{itemize}

\par\noindent
Second we write  Eq.(\ref{self-adjoint-eqn}) for the function $\psi_{-}$ and we get

\be
\left(\hat{p}^{\dagger} +\frac{i}{\alpha}\right)\,\psi_{-}(x) = 0 \notag \\
\ee
\par\noindent
which in terms of the dimensionless parameter $\rho$ takes the form
\be
\left(2\,\frac{d^{3}}{d\rho^{3}}-\frac{d}{d\rho}+1\right)\,\psi_{-}(\rho) = 0 \notag~.
\ee
\par\noindent
The characteristic equation of the above differential equation is
\be
2\,\mu^{3}-\mu+1 = 0
\ee
\par\noindent
and so its roots are of the form: $\mu^{1}_{-} = -1$, $\mu^{2}_{-} = +0.5 - 0.5\,i$, and $\mu^{3}_{-} = +0.5 + 0.5\,i$~. 
Therefore, we obtain three  linearly independent solutions for $\psi_{-}$ 
\bea
\psi_{-}^{1}(x) &\propto &  \exp\left(-\rho\right) \\
\psi_{-}^{2}(x) &\propto & \exp\left[\left(+0.5 - 0.5\,i\right)\rho\right]\\
\psi_{-}^{3}(x) &\propto &\exp\left[\left(+0.5 + 0.5\,i\right)\rho\right]~.
\eea
\par\noindent
Furthermore, over the different domains the deficiency index $n_{-}$  of the GUP-modified 
momentum operator which is quadratic in $\alpha$  is given as follows:

\begin{itemize}

\item $\mathcal{D}(\hat{p}) = \mathcal{L}^{2}(-\infty,\infty)$

In this domain, none of the functions, i.e., $\psi_{-}^{1}(x)$, $\psi_{-}^{2}(x)$, and $\psi_{-}^{3}(x)$, is square integrable. 
So these functions do not belong to this space and thus $n_{-} = 0$.\\

\item $\mathcal{D}(\hat{p}) = \mathcal{L}^{2}(0,\infty)$

In this domain, only $\psi_{-}^{1}(x)$ has finite norm. So this is the only solution from the set which is
square integrable and thus $n_{-} = 1$.\\

\item $\mathcal{D}(\hat{p}) = \mathcal{L}^{2}([a,b])$

Over the finite interval all three solutions, i.e., $\psi_{-}^{1}(x)$, $\psi_{-}^{2}(x)$, and $\psi_{-}^{3}$, are square integrable. 
So these functions belong to this space and thus $n_{+} = 3$.

\end{itemize}

\par\noindent
Finally, employing von Newmann's theorem,  the self-adjointness of the GUP-modified momentum operator which 
is quadratic in $\alpha$ is described as follows :

\begin{itemize}

\item $\mathcal{D}(\hat{p}) = \mathcal{L}^{2}(-\infty,\infty) : (n_{+},n_{-}) = (0,0)$ and thus the momentum operator is self-adjoint.

\item $\mathcal{D}(\hat{p}) = \mathcal{L}^{2}(0,\infty) : (n_{+},n_{-}) = (2,1)$ and thus the momentum operator is not self-adjoint.

\item $\mathcal{D}(\hat{p}) = \mathcal{L}^{2}([a,b]) : (n_{+},n_{-}) = (3,3)$ and thus the momentum operator has infinitely many self-adjoint extensions.

\end{itemize}

\subsection{Momentum operator with linear and quadratic terms in $\alpha$}
\par\noindent
We now consider as momentum operator in Eq.(\ref{self-adjoint-eqn}), the GUP-modified 
momentum operator which has linear and  quadratic terms in $\alpha$, namely
\be
\hat{p} = \hat{p}_{0}-\alpha\hat{p}_{0}^{2}+2\alpha^{2}\hat{p}_{0}^{3}~.
\ee
\par\noindent
First we write  Eq.(\ref{self-adjoint-eqn}) for the function $\psi_{+}$ and we get
\be
\left(\hat{p}^{\dagger} - \frac{i}{\alpha}\right)\,\psi_{+}(x) = 0 \notag 
\ee
\par\noindent
which in terms of the dimensionless parameter $\rho$ takes the form
\be
\left(2i\,\frac{d^{3}}{d\rho^{3}}+\frac{d^{2}}{d\rho^{2}}-i\,\frac{d}{d\rho}-i\right)\,\psi_{+}(\rho) = 0 \notag~.
\ee
\par\noindent
The characteristic equation of the above differential equation is
\be
2i\nu^{3}+\nu^{2}-i\nu-i = 0
\ee
\par\noindent
and so its roots are of the form: $ \nu^{1}_{+} = +1.0 + 0.2\, i$,  $\nu^{2}_{+} = -0.5 + 0.7 \,i$, and $\nu^{3}_{+} = 
- 0.4 - 0.4\,i$~.
Therefore, we obtain three  linearly independent solutions for $\psi_{+}$ 
\bea
\psi_{+}^{1}(\rho) &\propto & = \exp\left[(+1.0+ 0.2\, i)\rho\right] \\
\psi_{+}^{2}(\rho) &\propto & = \exp\left[(- 0.5 + 0.7\,i)\rho\right] \\
\psi_{+}^{3}(\rho) &\propto & = \exp\left[(- 0.4 - 0.4\,i)\rho\right]
\eea
\par\noindent
Furthermore, over the different domains the deficiency index $n_{+}$  of the GUP-modified 
momentum operator with linear and quadratic terms in $\alpha$  is given as follows:

\begin{itemize}

\item $\mathcal{D}(\hat{p}) = \mathcal{L}^{2}(-\infty, \infty)$

In this domain, none of the functions, i.e., $\psi_{+}^{1}(x)$, $\psi_{+}^{2}(x)$, and $\psi_{+}^{3}$, is square integrable. 
So these functions do not belong to this space and thus $n_{+} = 0$.\\

\item $\mathcal{D}(\hat{p}) = \mathcal{L}^{2}(0,\infty)$

In this domain, only $\psi_{+}^{2}(x)$ and $\psi_{+}^{3}$ have finite norm. So there are only two  solutions 
from the set which are square integrable and thus $n_{+} = 2$.\\

\item $\mathcal{D}(\hat{p}) = \mathcal{L}^{2}([a,b])$

Over the finite interval all three solutions, i.e., $\psi_{+}^{1}(x)$, $\psi_{+}^{2}(x)$, and $\psi_{+}^{3}$, are square integrable. 
So these functions belong to this space and thus $n_{+} = 3$.

\end{itemize}
\par\noindent
Second we write  Eq.(\ref{self-adjoint-eqn}) for the function $\psi_{-}$ and we get
\be
\left(\hat{p}^{\dagger} + \frac{i}{\alpha}\right)\,\psi_{-}(\rho) = 0 \notag 
\ee
\par\noindent
which in terms of the dimensionless parameter $\rho$ takes the form
\be
\left(2i\,\frac{d^{3}}{d\rho^{3}}+\frac{d^{2}}{d\rho^{2}}-i\,\frac{d}{d\rho}+i\right)\,\psi_{+}(\rho) = 0 \notag~.
\ee
\par\noindent
The characteristic equation of the above differential equation is
 \be
2i\nu^{3}+\nu^{2}-i\nu+i = 0
\ee
\par\noindent
and so its roots are of the form:  $\nu^{1}_{-} =+ 0.5 + 0.7\,i$, $\nu^{2}_{-} =+0.4 - 0.4\,i$, and $\nu^{3}_{-} = -1.0 
+ 0.2\,i$~.
Therefore, we obtain three  linearly independent solutions for $\psi_{+}$ 
\bea
\psi_{-}^{4}(\rho) &\propto & = \exp\left[(+0.5+ 0.7\,i)\rho\right] \\
\psi_{-}^{5}(\rho) &\propto & = \exp\left[(+0.4 - 0.4\,i)\rho\right] \\
\psi_{-}^{6}(\rho) &\propto & = \exp\left[(- 1.0+ 0.2\,i)\rho\right]
\eea
\par\noindent
Furthermore, over the different domains the deficiency index $n_{-}$  of the GUP-modified 
momentum operator with linear and quadratic terms in  $\alpha$  is given as follows:

\begin{itemize}

\item $\mathcal{D}(\hat{p}) = \mathcal{L}^{2}(-\infty, \infty)$

In this domain, none of the functions, i.e., $\psi_{-}^{4}(x)$, $\psi_{-}^{5}(x)$, and $\psi_{-}^{6}$, is square integrable. 
So these functions do not belong to this space and thus $n_{-} = 0$.\\

\item $\mathcal{D}(\hat{p}) = \mathcal{L}^{2}(0,\infty)$

In this domain, only $\psi_{-}^{6}(x)$ has finite norm. So this is the only solution from the set which is
square integrable and thus $n_{-} = 1$.\\

\item $\mathcal{D}(\hat{p}) = \mathcal{L}^{2}([a,b]),$

Over the finite interval all three solutions, i.e., $\psi_{-}^{4}(x)$, $\psi_{-}^{5}(x)$, and $\psi_{-}^{3}$, are square integrable. 
So these functions belong to this space and thus $n_{-} = 3$.

\end{itemize}

\par\noindent
Finally, employing von Newmann's theorem,  the self-adjointness of the GUP-modified momentum operator which 
is quadratic in $\alpha$ is described as follows:

\begin{itemize}

\item $\mathcal{D}(\hat{p}) = \mathcal{L}^{2}(-\infty, \infty) :(n_{+}, n_{-}) = (0,0)$ and thus the momentum operator is self-adjoint.

\item $\mathcal{D}(\hat{p}) = \mathcal{L}^{2}(0, \infty) : (n_{+},n_{-}) = (2,1)$ and thus the momentum operator is not self-adjoint.

\item $\mathcal{D}(\hat{p}) = \mathcal{L}^{2}([a,b]) : (n_{+},n_{-}) = (3,3)$  and thus the momentum operator has 
infinitely many self-adjoint extensions.

\end{itemize}

\par\noindent
All results produced in this section are briefly presented in Table I
(all A's, B's, C's, and D's are constants).
\begin{table}[h!b!p!]\large
\caption{Results for GUP-modified Momentum operator}
    \begin{tabular}[t]{*{6}{|c}|}
            \hline
 &  &  &   & $(n_{+},n_{-})$ &   \\ \cline{4-6}
\raisebox{2.0ex}[0pt]{\textbf{Operator}}&\raisebox{2.0ex}[0pt]{\textbf{$\psi_{+}(\rho)$}}&\raisebox{2.0ex}[0pt]{\textbf{$\psi_{-}(\rho)$}}& $(-\infty, +\infty)$ & $[0,+ \infty)$ & $[a,b]$  \\\cline{1-6}
$\hat{p}_{0}$ & $A_{1}\exp\left[-\rho\right]$ & $A_{2}\exp\left[\rho\right]$  & $(0,0)$ & $(1,0)$ & $(1,1)$   \\\cline{1-6}

& $B_{1}\exp\left[\lambda^{1}_{+}\rho\right]$  & $B_{3}\exp\left[\lambda^{1}_{-}\rho\right]$ &  &  &   \\\cline{2-3}
\raisebox{2.0ex}[0pt]{$\hat{p}_{0}(1-\alpha\hat{p}_{0})$}   & $B_{2}\exp\left[\lambda^{2}_{+}\rho\right]$ & $B_{4}\exp\left[\lambda^{2}_{-}\rho\right]$ &\raisebox{2.0ex}[0pt]{$(0,0)$}  &\raisebox{2.0ex}[0pt]{$(1,1)$}  & \raisebox{2.0ex}[0pt]{$(2,2)$}    \\\cline{1-6}

 & $C_{1}\exp\left[\mu^{1}_{+}\rho\right]$ & $C_{4}\exp\left[\mu^{1}_{-}\rho\right]$  & & &  \\\cline{2-3}
 $\hat{p}_{0}\left(1+2\alpha^{2}\hat{p}^{2}_{0}\right)$   & $C_{2}\exp\left[\mu^{2}_{+}\rho\right]$ & $C_{5}\exp\left[\mu^{2}_{-}\rho\right]$  &  $(0,0)$ &  $(2,1)$ & $(3,3)$  \\\cline{2-3}
            & $C_{3}\exp\left[\mu^{3}_{+}\rho\right]$ & $C_{6}\exp\left[\mu^{3}_{-}\rho\right]$  & & &  \\ \cline{1-6}

 & $D_{1}\exp\left[\nu^{1}_{+}\rho\right]$ & $D_{4}\exp\left[\nu^{1}_{-}\rho\right]$  & & &  \\\cline{2-3}
    $\hat{p}_{0}\left(1-\alpha\hat{p}^{2}_{0}+2\alpha^{2}\hat{p}^{3}_{0}\right)$ & $D_{2}\exp\left[\nu^{2}_{+}\rho\right]$ & $D_{5}\exp\left[\nu^{2}_{-}\rho\right]$ & $(0,0)$  & $(2,1)$ &  $(3,3)$ \\\cline{2-3}
           & $D_{3}\exp\left[\nu^{3}_{+}\rho\right]$ & $D_{6}\exp\left[\nu^{6}_{-}\rho\right]$  & & &   \\ \cline{1-6}
      \end{tabular}
\end{table}

\section{GUP-modified Hamiltonian}

\par\noindent
In this section we will apply von Neumann's theorem to the GUP-modified Hamiltonian operator. For this reason, 
we first have to determine the functions $\psi_{\pm}(x)$ \cite{Bonneau:1999zq} which satisfy the equation
\be
\hat{H} \psi_{\pm}(x) = \pm ik_{0}^{2}\psi_{\pm}(x)
\label{hamiltonian_self-adjoint-eqn}
\ee
\par\noindent
where $k_{0}$ is a  positive constant.\\
For the  simple case of the Hamiltonian of  a free particle, i.e., $\hat{H} = -D^{2}$, where $D$ 
is the differential $d/dx$ in the standard representation,  Eq.(\ref{hamiltonian_self-adjoint-eqn}) becomes
\be
  -D^{2}\psi_{\pm}(x) = \pm ik_{0}^{2}\psi_{\pm}(x)~.
\ee
\par\noindent
It is easily seen that the linearly independent solutions to the above equation are of the form
\be
\psi_{\pm}(x) = a_{\pm}\exp\left[k_{\pm}x\right]+b_{\pm}\exp\left[-k_{\pm}x\right]
\ee
\par\noindent
where $k_{\pm} = \frac{(1\mp i)}{\sqrt{2}}k_{0}$. 

\par\noindent
Over different domains, the deficiency indices and the self-adjointness of the Hamiltonian operator are described as follows:

\begin{itemize}

\item $\mathcal{D}(\hat{H}_{0}) = \mathcal{L}^{2}(\mathbb{R})$ :  $(n_{+},n_{-})=(0,0)$ 
and thus  the operator is self-adjoint.

\item $\mathcal{D}(\hat{H}_{0}) = \mathcal{L}^{2}[0,\infty)$ :  $(n_{+},n_{-})=(1,1)$ 
and thus the operator has infinitely many self-adjoint
extensions parametrized by a $U(1)$ group.

\item $\mathcal{D}(\hat{H}_{0}) = \mathcal{L}^{2}([a,b])$:  $(n_{+},n_{-})=(2,2)$
and thus the operator has infinitely many self-adjoint
extensions parametrized by a $U(2)$ group.

\end{itemize}

\par\noindent
At this point it is noteworthy that since the quantity $(m \alpha^{2})^{-1}$ (with $m$ to be the mass of the particle) 
is a characteristic energy scale of GUP,  we will use it from now on instead of $k_{0}$.

\subsection{Hamiltonian with linear term in $\alpha$}
\par\noindent
We now consider as momentum operator in Eq.(\ref{hamiltonian_self-adjoint-eqn}), the GUP-modified 
Hamiltonian operator which is linear in $\alpha$, namely
\be
\hat{H} = \frac{\hat{p}_{0}^{2}}{2m}-\frac{\alpha}{m}\hat{p}_{0}^{3}
\ee
\par\noindent
and we adopt the analysis of the previous section in order to study the self-adjointness of this GUP-modified 
Hamiltonian operator.

\par\noindent
First we write  Eq.(\ref{hamiltonian_self-adjoint-eqn}) for the function $\psi_{+}$ and we get
\be
\left(\hat{H}^{\dagger}-\frac{i}{\alpha^{2}m}\right)\psi_{+}(x) = 0 \notag 
\ee
\par\noindent
which in terms of the dimensionless parameter $\rho$ takes the form
\be
\left(2i\,\frac{d^{3}}{d\rho^{3}}+\frac{d^{2}}{d\rho^{2}}+2i\right)\,\psi_{+}(\rho) = 0 \notag~.
\ee
\par\noindent
The characteristic equation of the above differential equation is
\be
2i\,\lambda^{3}+\lambda^{2}+2i = 0
\ee
\par\noindent
and so its roots are of the form:  $\lambda^{1}_{+} =- 1.0 + 0.2\,i$, $\lambda^{2}_{+} =+0.5 + 1.1\,i$, and $\lambda^{3}_{+} 
=+ 0.5- 0.7\,i$~. 
Therefore, we obtain three  linearly independent solutions for $\psi_{+}$ 
\bea
\psi_{+}^{1}(\rho) &\propto & \exp\left[(- 1.0 + 0.2\,i)\rho\right] \\
\psi_{+}^{2}(\rho) &\propto & \exp\left[(+0.5 + 1.1\,i)\rho\right] \\
\psi_{+}^{3}(\rho) &\propto & \exp\left[(+0.5 - 0.7\,i)\rho\right]~.
\eea
\par\noindent
Furthermore, over the different domains, the deficiency index $n_{+}$  of the GUP-modified 
Hamiltonian  operator which is  linear in $\alpha$  is given as follows:

\begin{itemize}

\item $\mathcal{D}(\hat{H}) = \mathcal{L}^{2}(-\infty, \infty)$

In this domain, none of the functions, i.e., $\psi_{+}^{1}(x)$, $\psi_{+}^{2}(x)$, and $\psi_{+}^{3}$, is square integrable. 
So these functions do not belong to this space and thus $n_{+} = 0$.\\

\item $\mathcal{D}(\hat{H}) = \mathcal{L}^{2}[0,\infty)$

In this domain, only $\psi_{+}^{1}(x)$  has finite norm. So this is the only solution from the set which is
square integrable and thus $n_{+} = 1$.\\

\item  $\mathcal{D}(\hat{H}) = \mathcal{L}^{2}([a,b])$

Over the finite interval all three solutions, i.e., $\psi_{+}^{1}(x)$, $\psi_{+}^{2}(x)$, and $\psi_{+}^{3}$, 
are square integrable.  So these functions belong to this space and thus $n_{+} = 3$.

 \end{itemize}

\par\noindent
Second we write  Eq.(\ref{self-adjoint-eqn}) for the function $\psi_{-}$ and we get
\be
\left(\hat{H}^{\dagger}+\frac{i}{\alpha^{2}m}\right)\psi_{-}(x) = 0 \notag 
\ee
\par\noindent
which in terms of the dimensionless parameter $\rho$ takes the form
\be
\left(2i\,\frac{d^{3}}{d\rho^{3}}+\frac{d^{2}}{d\rho^{2}}-2i\right)\,\psi_{-}(\rho) = 0 \notag~.
\ee

\par\noindent
The characteristic equation of the above differential equation is
\be
2i\,\lambda^{3}+\lambda^{2}-2i = 0
\ee
\par\noindent
and so its roots are of the form: $\lambda^{1}_{-}  = +1.0 + 0.2\,i$, $\lambda^{2}_{-} =- 0.5 + 1.1\,i$, and $\lambda^{3}_{-} 
=- 0.5 - 0.7\,i$~.
Therefore, we obtain three  linearly independent solutions for $\psi_{-}$ 
\bea
\psi_{-}^{1}(\rho) &\propto &\exp\left[(+1.0 + 0.2\,i)\rho\right] \\
\psi_{-}^{2}(\rho) &\propto & \exp\left[(- 0.5+ 1.1\,i)\rho\right] \\
\psi_{-}^{3}(\rho) &\propto & \exp\left[(- 0.5- 0.7\,i)\rho\right]~.
\eea
\par\noindent
Furthermore, over the different domains the deficiency index $n_{-}$  of the GUP-modified 
Hamiltonian operator which is linear in $\alpha$  is given as follows:

\begin{itemize}

\item $\mathcal{D}(\hat{H}) = \mathcal{L}^{2}(-\infty,\infty)$

In this domain, none of the functions, i.e., $\psi_{-}^{1}(x)$, $\psi_{-}^{2}(x)$, and $\psi_{-}^{3}(x)$, is square integrable. 
So these functions do not belong to this space and thus $n_{-} = 0$.\\

\item $\mathcal{D}(\hat{H}) = \mathcal{L}^{2}[0,\infty)$

In this domain, only $\psi_{-}^{2}(x)$ and $\psi_{-}^{3}(x)$  have finite norm. So there are only two solutions 
from the set which are square integrable and thus $n_{-} = 2$.\\

\item $\mathcal{D}(\hat{H}) = \mathcal{L}^{2}([a,b])$

Over the finite interval both solutions, i.e., $\psi_{-}^{1}(x)$, $\psi_{-}^{2}(x)$, and $\psi_{-}^{3}$, are square integrable. 
So these functions belong to this space and thus $n_{-} = 3$.

\end{itemize}

\par\noindent
Finally, employing von Newmann's theorem,  the self-adjointness of the GUP-modified Hamiltonian operator which 
is linear in $\alpha$ is described as follows :

\begin{itemize}

\item $\mathcal{D}(\hat{H}) = \mathcal{L}^{2}(-\infty,\infty) : (n_{+},n_{-}) = (0,0)$. and thus the Hamiltonian operator is self-adjoint.

\item $\mathcal{D}(\hat{H}) = \mathcal{L}^{2}(0,\infty) : (n_{+},n_{-}) = (1,2)$ and thus the Hamiltonian operator is not self-adjoint.

\item $\mathcal{D}(\hat{H}) = \mathcal{L}^{2}([a,b]) : (n_{+},n_{-}) = (3,3)$ and thus the Hamiltonian operator has infinitely many
 self-adjoint extensions parametrized by a $U(3)$ group.

 \end{itemize}

\subsection{Hamiltonian with quadratic term in $\alpha$}
\par\noindent
Let us now consider as momentum operator in Eq.(\ref{hamiltonian_self-adjoint-eqn}), the GUP-modified 
Hamiltonian operator which is quadratic in $\alpha$, namely
\bea
\hat{H} &=& \frac{\hat{p}_{0}^{2}}{2m}+\frac{5\alpha^{2}}{2m}\hat{p}_{0}^{4}\notag \\
\eea
\par\noindent
First we write  Eq.(\ref{hamiltonian_self-adjoint-eqn}) for the function $\psi_{+}$ and we get
\be
\left(\hat{H}^{\dagger}-\frac{i}{\alpha^{2}m}\right)\,\psi_{+}(x) = 0 \notag 
\ee
\par\noindent
which in terms of the dimensionless parameter $\rho$ takes the form
\be
\left(5\,\frac{d^{4}}{d\rho^{4}}-\frac{d^{2}}{d\rho^{2}}-2\,i\right)\,\psi_{+}(\rho) = 0 \notag~.
\ee
\par\noindent
The characteristic equation of the above differential equation is
\be
5\mu^{4}-\mu^{2}-2i=0
\ee
\par\noindent
and so its roots are of the form: $\mu^{1}_{+} = +0.8 + 0.3\,i$, $\mu^{2}_{+} = - 0.8 - 0.3\,i$, 
$\mu^{3}_{+} = +0.3 - 0.7\,i$, and $\mu^{4}_{+} = - 0.3 + 0.7\,i$~.
Therefore, we obtain four  linearly independent solutions for $\psi_{+}$ 
\bea
\psi_{+}^{1}(\rho)  &\propto & \exp\left[(+0.8+ 0.3\,i)\rho\right] \notag \\
\psi_{+}^{2}(\rho) &\propto & \exp\left[(- 0.8 - 0.3\,i)\rho\right] \notag \\
\psi_{+}^{3}(\rho)  &\propto & \exp\left[(+0.3 - 0.7\,i)\rho\right] \notag \\
\psi_{+}^{4}(\rho)  &\propto & \exp\left[(- 0.3 + 0.7\,i)\rho\right]\notag~.
\eea
\par\noindent
Furthermore, over the different domains, the deficiency index $n_{+}$  of the GUP-modified 
Hamiltonian  operator which is  quadratic in $\alpha$  is given as follows:

\begin{itemize}

\item $\mathcal{D}(\hat{H}) = \mathcal{L}^{2}(\mathbb{R})$

In this domain, none of the functions, i.e., $\psi_{+}^{1}(x)$, $\psi_{+}^{2}(x)$, $\psi_{+}^{3}$, and $\psi_{+}^{4}$, is square integrable. 
So these functions do not belong to this space and thus $n_{+} = 0$.\\

\item $\mathcal{D}(\hat{H}) = \mathcal{L}^{2}[0,\infty)$

In this domain, only $\psi_{+}^{2}(x)$ and $\psi^{4}_{+}$  have finite norm. So there are only two solutions from the set which is
square integrable and thus $n_{+} = 2$.\\

\item $\mathcal{D}(\hat{H}) = \mathcal{L}^{2}([a,b])$

Over the finite interval all four solutions, i.e., $\psi_{+}^{1}(x)$, $\psi_{+}^{2}(x)$, $\psi_{+}^{3}$, and $\psi_{+}^{4}$, 
are square integrable.  So these functions belong to this space and thus $n_{+} = 4$.

\end{itemize}

\par\noindent
Second we write  Eq.(\ref{self-adjoint-eqn}) for the function $\psi_{-}$ and we get
\be
\left(\hat{H}^{\dagger}+\frac{i}{\alpha^{2}m}\right)\,\psi_{-}(\rho) = 0 \notag 
\ee
\par\noindent
which in terms of the dimensionless parameter $\rho$ takes the form
\be
\left(5\,\frac{d^{4}}{d\rho^{4}}-\frac{d^{2}}{d\rho^{2}}+2i\right)\,\psi_{-}(\rho) = 0 \notag~.
\ee
\par\noindent
The characteristic equation of the above differential equation is
\be
5\,\mu^{4}-\mu^{2}+2i = 0
\ee
\par\noindent
and so its roots are of the form: $\mu^{1}_{-}  = +0.8 - 0.3\,i$, 
$\mu^{2}_{-} =- 0.8 + 0.3\,i$, $\mu^{3}_{-} = 0.3 + 0.7\,i$, and $\mu^{4}_{-} = - 0.3 - 0.7\,i$~. 
Therefore, we obtain four  linearly independent solutions for $\psi_{-}$ 
\bea
\psi_{-}^{1}(\rho) &\propto &\exp\left[(+0.8- 0.3\,i)\rho\right] \notag \\
\psi_{-}^{2}(\rho) &\propto &\exp\left[(- 0.8 + 0.3\,i)\rho\right] \notag \\
\psi_{-}^{3}(\rho) &\propto & \exp\left[(+0.3 + 0.7\,i)\rho\right] \notag \\
\psi_{-}^{4}(\rho) &\propto &\exp\left[(- 0.3 - 0.7\,i)\rho\right] \notag~.
\eea
\par\noindent
Furthermore, over the different domains the deficiency index $n_{-}$  of the GUP-modified 
Hamiltonian operator which is quadratic in $\alpha$  is given as follows:

\begin{itemize}

\item $\mathcal{D}(\hat{H}) = \mathcal{L}^{2}(-\infty, \infty)$

In this domain, none of the functions, i.e., $\psi_{-}^{1}(x)$, $\psi_{-}^{2}(x)$, $\psi_{-}^{3}(x)$, and $\psi^{4}_{-}$, is square integrable. 
So these functions do not belong to this space and thus $n_{-} = 0$.\\

\item $\mathcal{D}(\hat{H}) = \mathcal{L}^{2}[0,\infty)$

In this domain, only  $\psi_{-}^{2}(x)$ and $\psi^{4}_{-}$  have finite norm. So there are only two solutions from the set which is
square integrable and thus $n_{-} = 2$.\\

\item $\mathcal{D}(\hat{H}) = \mathcal{L}^{2}([a,b])$

Over the finite interval all four solutions, i.e., $\psi_{-}^{1}(x)$, $\psi_{-}^{2}(x)$, $\psi_{-}^{3}$, and $\psi_{-}^{4}$, 
are square integrable.  So these functions belong to this space and thus $n_{-} = 4$.

\end{itemize}

\par\noindent
Finally, employing von Newmann's theorem,  the self-adjointness of the GUP-modified Hamiltonian operator which 
is quadratic in $\alpha$ is described as follows :

\begin{itemize}

\item $\mathcal{D}(\hat{H}) = \mathcal{L}^{2}(-\infty,\infty) : (n_{+},n_{-}) = (0,0)$ and thus the Hamiltonian operator is self-adjoint.

\item $\mathcal{D}(\hat{H}) = \mathcal{L}^{2}(0,\infty) : (n_{+},n_{-}) = (2,2)$ and thus the Hamiltonian operator has infinitely many
 self-adjoint extensions parametrized by a $U(2)$ group.

\item $\mathcal{D}(\hat{H}) = \mathcal{L}^{2}([a,b]) : (n_{+},n_{-}) = (4,4)$ and thus the momentum operator has infinitely many
 self-adjoint extensions parametrized by a $U(4)$ group.

\end{itemize}

\subsection{Hamiltonian operator with linear and quadratic terms in $\alpha$}
\par\noindent
We now consider as momentum operator in Eq.(\ref{hamiltonian_self-adjoint-eqn}), the GUP-modified 
Hamiltonian operator which has linear and quadratic terms in $\alpha$, namely
\be
\hat{H} = \frac{\hat{p}_{0}^{2}}{2m}-\frac{\alpha}{m}\hat{p}_{0}^{3}+\frac{5\alpha^{2}}{2m}\hat{p}_{0}^{4}~.
\ee
\par\noindent
First we write  Eq.(\ref{hamiltonian_self-adjoint-eqn}) for the function $\psi_{+}$ and we get
\be
\left(\hat{H}^{\dagger}-\frac{i}{\alpha^{2}m}\right)\,\psi_{+}(x) = 0 \notag 
\ee
\par\noindent
which in terms of the dimensionless parameter $\rho$ takes the form
\be
\left(5\,\frac{d^{4}}{d\rho^{4}}-2i\,\frac{d^{3}}{d\rho^{3}}-\frac{d^{2}}{d\rho^{2}}-2i\right)\,\psi_{+}(\rho) = 0 \notag~.
\ee
\par\noindent
The characteristic equation of the above differential equation is
\be
5\,\nu^{4}-2i\,\nu^{3}-\nu^{2}-2i = 0~.
\ee
\par\noindent
Numerically solving the above characteristic equation, its roots are of the form: 
$\nu^{1}_{+}  = - 0.8 - 0.2\,i$, $\nu^{2}_{+} = - 0.3 + 0.8\,i$, $\nu^{3}_{+} = + 0.3 - 0.6\,i$, and 
$\nu^{4}_{+} = + 0.8+ 0.4\,i$~. 
Therefore, we obtain four  linearly independent solutions for $\psi_{+}$ 
\bea
\psi_{+}^{1}(\rho) &\propto &\exp\left[( - 0.8 - 0.2\,i )\rho\right] \notag \\
\psi_{+}^{2}(\rho) &\propto &\exp\left[( - 0.3 + 0.8\,i)\rho\right] \notag \\
\psi_{+}^{3}(\rho) &\propto & \exp\left[(+0.3 - 0.6\,i)\rho\right] \notag \\
\psi_{+}^{4}(\rho) &\propto &\exp\left[(+0.8 + 0.4\,i)\rho\right] \notag~.
\eea

\par\noindent
Furthermore, over the different domains, the deficiency index $n_{+}$  of the GUP-modified 
Hamiltonian  operator which has  linear and quadratic terms in $\alpha$  is given as follows:

\begin{itemize}

\item $\mathcal{D}(\hat{H}) = \mathcal{L}^{2}(-\infty, \infty)$

In this domain, none of the functions, i.e., $\psi_{+}^{1}(x)$, $\psi_{+}^{2}(x)$, $\psi_{+}^{3}$, and $\psi_{+}^{4}$, is square integrable. 
So these functions do not belong to this space and thus $n_{+} = 0$.\\

\item $\mathcal{D}(\hat{H}) = \mathcal{L}^{2}[0,\infty)$

In this domain, only $\psi_{+}^{1}(x)$ and $\psi^{2}_{+}$  have finite norm. So there are only two solutions from the set which are
square integrable and thus $n_{+} = 2$.\\

\item  $\mathcal{D}(\hat{H}) = \mathcal{L}^{2}([a,b])$

Over the finite interval all four solutions, i.e., $\psi_{+}^{1}(x)$, $\psi_{+}^{2}(x)$,  $\psi_{+}^{3}$,  and $\psi^{4}_{+}$,
are square integrable.  So these functions belong to this space and thus $n_{+} = 4$.

\end{itemize}

\par\noindent
Second we write  Eq.(\ref{hamiltonian_self-adjoint-eqn}) for the function $\psi_{-}$ and we get
\be
\left(\hat{H}^{\dagger}+\frac{i}{\alpha^{2}m}\right)\,\psi_{+}(x) = 0 \notag 
\ee
\par\noindent
which in terms of the dimensionless parameter $\rho$ takes the form
\be
\left(5\,\frac{d^{4}}{d\rho^{4}}-2i\,\frac{d^{3}}{d\rho^{3}}-\frac{d^{2}}{d\rho^{2}}+2i\right)\,\psi_{+}(\rho) = 0 \notag~.
\ee
\par\noindent
The characteristic equation of the above differential equation is
\bea
5\,\lambda^{4}-2i\,\lambda^{3}-\lambda^{2}+2i = 0~.
\eea

\par\noindent
Numerically solving the above characteristic equation, its roots are of the form: 
$\nu^{1}_{-} = - 0.8 + 0.4\,i$, $\nu^{2}_{-} = - 0.3 - 0.6\,i$, $\nu^{3}_{-} = + 0.3+ 0.8\,i$, and $\nu^{4}_{-} =+ 0.8- 0.2\,i$~.
Therefore, we obtain four  linearly independent solutions for $\psi_{-}$
\bea
\psi_{-}^{1}(\rho) &\propto &\exp\left[(- 0.8 + 0.4\,i )\rho\right] \notag \\
\psi_{-}^{2}(\rho) &\propto &\exp\left[( - 0.3- 0.6\,i)\rho\right] \notag \\
\psi_{-}^{3}(\rho) &\propto & \exp\left[( +0.3 + 0.8\,i)\rho\right] \notag \\
\psi_{-}^{4}(\rho) &\propto &\exp\left[( +0.8 - 0.2\,i)\rho\right] \notag~.
\eea

\par\noindent
Furthermore, over the different domains, the deficiency index $n_{+}$  of the GUP-modified 
Hamiltonian  operator which has  linear and quadratic terms in $\alpha$  is given as follows:

\begin{itemize}

\item $\mathcal{D}(\hat{H}) = \mathcal{L}^{2}(-\infty, \infty)$

In this domain, none of the functions, i.e., $\psi_{-}^{1}(x)$, $\psi_{-}^{2}(x)$, $\psi_{-}^{3}$, and $\psi_{-}^{4}$, is square integrable. 
So these functions do not belong to this space and thus $n_{-} = 0$.\\

\item $\mathcal{D}(\hat{H}) = \mathcal{L}^{2}[0,\infty)$

In this domain, only $\psi_{-}^{1}(x)$ and $\psi^{2}_{-}$  have finite norm. So there are only two solutions from the set which are
square integrable and thus $n_{-} = 2$.\\

\item  $\mathcal{D}(\hat{H}) = \mathcal{L}^{2}([a,b])$

Over the finite interval all four solutions, i.e., $\psi_{-}^{1}(x)$, $\psi_{-}^{2}(x)$,  $\psi_{-}^{3}$,  and $\psi^{4}_{-}$,
are square integrable.  So these functions belong to this space and thus $n_{-} = 4$.

\end{itemize}

\par\noindent
Finally, employing von Newmann's theorem,  the self-adjointness of the GUP-modified Hamiltonian operator which 
is linear in $\alpha$ is described as follows :

\begin{itemize}

\item $\mathcal{D}(\hat{H}) = \mathcal{L}^{2}(-\infty,\infty) : (n_{+},n_{-}) = (0,0)$. and thus the Hamiltonian operator is self-adjoint.

\item $\mathcal{D}(\hat{H}) = \mathcal{L}^{2}(0,\infty) : (n_{+},n_{-}) = (2,2)$ and thus the Hamiltonian  operator has infinitely many
 self-adjoint extensions parametrized by a $U(2)$ group.

\item $\mathcal{D}(\hat{H}) = \mathcal{L}^{2}([a,b]) : (n_{+},n_{-}) = (4,4)$ and thus the Hamiltonian  operator has infinitely many
 self-adjoint extensions parametrized by a $U(4)$ group.

 \end{itemize}

\par\noindent
All results produced in this section are briefly presented in Table II  (all $\tilde{A}$'s, $\tilde{B}$'s, $\tilde{C}$'s, and $\tilde{D}$'s are constants).
\begin{table}[h!b!p!]\large
\caption{Results for GUP-modified Hamiltonian operator}
    \begin{tabular}[t]{*{7}{|c}|}

\hline
 &  & & & $(n_{+},n_{-})$     &        \\ \cline{4-6}
 \raisebox{2.0ex}[0pt]{\textbf{Operator}}& \raisebox{2.0ex}[0pt]{\textbf{$\psi_{+}(\rho)$}}& \raisebox{2.0ex}[0pt]{\textbf{$\psi_{-}(\rho)$}}  & $(-\infty,\infty)$ & $[0,\infty)$ & $[a,b]$  \\ \cline{1-6}

& $\tilde{A}_{1}\exp\left[+k_{+}x\right]$  &  $\tilde{A}_{3}\exp\left[+k_{-}x\right]$  &  & &  \\ \cline{2-3}
  \raisebox{2.0ex}[0pt]{$\frac{1}{2m}\hat{p}^{2}_{0}$}      &  $\tilde{A}_{2}\exp\left[-k_{+}x\right]$ & $\tilde{A}_{4}\exp\left[-k_{-}x\right]$  &  \raisebox{2.0ex}[0pt]{$(0,0)$ }  &  \raisebox{2.0ex}[0pt]{$(1,1)$ }  &  \raisebox{2.0ex}[0pt]{$(2,2)$} \\ \cline{1-6}

   & $\tilde{B_{1}}\exp\left[\lambda^{1}_{+}\rho\right]$  & $\tilde{B_{4}}\exp\left[\lambda^{1}_{-}\rho\right]$ & &  &  \\ \cline{2-3}
$\frac{1}{2m}\hat{p}^{2}_{0}-\frac{\alpha}{m}\hat{p}_{0}^{3}$ & $\tilde{B_{2}}\exp\left[\lambda^{2}_{+}\rho\right]$ & $\tilde{B_{5}}\exp\left[\lambda^{2}_{-}\rho\right]$  &  $(0,0)$  & $(1,2)$   & $(3,3)$   \\ \cline{2-3}
                  & $\tilde{B_{3}}\exp\left[\lambda^{3}_{+}\rho\right]$ & $\tilde{B_{6}}\exp\left[\lambda^{3}_{-}\rho\right]$  &  & & \\ \cline{1-6}

& $\tilde{C_{1}}\exp\left[\mu^{1}_{+}\rho\right]$   & $\tilde{C_{5}}\exp\left[\mu^{1}_{-}\rho\right]$   & & &  \\ \cline{2-3}
 & $\tilde{C_{2}}\exp\left[\mu^{2}_{+}\rho\right]$  & $\tilde{C_{6}}\exp\left[\mu^{2}_{-}\rho\right]$  & & &  \\ \cline{2-3}
\raisebox{2.0ex}[0pt]{$\frac{1}{2m}\hat{p}^{2}_{0}+\frac{5\alpha^{2}}{2m}\hat{p}^{4}_{0}$}   &   $\tilde{C_{3}}\exp\left[\mu^{3}_{+}\rho\right]$   &  $\tilde{C_{7}}\exp\left[\mu^{3}_{-}\rho\right]$   &  \raisebox{2.0ex}[0pt]{$(0,0)$}&  \raisebox{2.0ex}[0pt]{$(2,2)$} &  \raisebox{2.0ex}[0pt]{$(4,4)$}   \\ \cline{2-3}
&  $\tilde{C_{4}}\exp\left[\mu^{4}_{+}\rho\right]$ &  $\tilde{C_{8}}\exp\left[\mu^{4}_{-}\rho\right]$   &  &  &  \\ \cline{1-6}

  & $\tilde{D_{1}}\exp\left[\nu^{1}_{+}\rho\right]$   & $\tilde{D_{5}}\exp\left[\nu^{1}_{-}\rho\right]$ & & &  \\ \cline{2-3}
  & $\tilde{D_{2}}\exp\left[\nu^{2}_{+}\rho\right]$  & $\tilde{D_{6}}\exp\left[\nu^{2}_{-}\rho\right]$ & & &  \\ \cline{2-3}
 \raisebox{2.0ex}[0pt]{$\frac{\hat{p}_{0}^{2}}{2m}-\frac{\alpha}{m}\hat{p}_{0}^{3}+\frac{5\alpha^{2}}{2m}\hat{p}_{0}^{4}$}    & $\tilde{D_{3}}\exp\left[\nu^{3}_{+}\rho\right]$  &  $\tilde{D_{8}}\exp\left[\nu^{3}_{-}\rho\right]$  &  \raisebox{2.0ex}[0pt]{ $(0,0)$} &  \raisebox{2.0ex}[0pt]{ $(2,2)$} &  \raisebox{2.0ex}[0pt]{$(4,4)$} \\ \cline{2-3}
  &$\tilde{D_{4}}\exp\left[\nu^{4}_{+}\rho\right]$  & $\tilde{D_{8}}\exp\left[\nu^{4}_{-}\rho\right]$  & & &  \\ \cline{1-6}

      \end{tabular}
\end{table}
\section{A simple Example}

\par\noindent
In this section we present a simple example of the GUP-modified Hamiltonian with a linear term in $\alpha$ \cite{Ali:2009zq}
\bea
H = \frac{p_0^2}{2m} - \frac{\alpha}{m} p_0^3~.
\eea
\par\noindent
We follow the analysis of \cite{Bonneau:1999zq} in order to describe the self-adjoint extensions of the specific Hamiltonian.
We choose the domain of the operator to be a finite interval of the positive semi-axis. 
Therefore, we obtain (we use ${}^\star$ to denote complex conjugates)\\
\bea
(H^\dagger \phi,\psi) - (\phi, H^\dagger \psi)  &=&\int_0^L \le[
\le( -\frac{\hbar^2}{2m} \frac{d^2\phi^\star}{dx^2} + \frac{i\alpha\hbar^3}{m} \frac{d^3 \phi^\star}{dx^3}
\ri) \psi  -\phi^\star \le( -\frac{\hbar^2}{2m} \frac{d^2\psi }{dx^2} - \frac{i\alpha\hbar^3}{m} \frac{d^3 \psi}{dx^3}
\ri)\ri] dx \nn \\
&=& \int_0^L \le[
\frac{\hbar^2}{2m} \le(
\phi^\star \frac{d^2\psi}{dx^2} - \psi \frac{d^2\phi^\star}{dx^2} \ri) 
%
\le(\frac{d^3\phi^\star}{dx^3} \psi + \phi^\star \frac{d^3\psi}{dx^3}\ri)\ri] dx \nn \\
&=& \int_0^L \le[
\frac{\hbar^2}{2m} \frac{d}{dx}
\le(\phi^\star \frac{d\psi}{dx} - \psi \frac{d\phi^\star}{dx}
\ri) +
\frac{i\alpha\hbar^3}{m}
\le(
\frac{d^2}{dx^2} \le( \phi^\star\psi \ri) - 3\frac{d\phi^\star}{dx} \frac{d\psi}{dx}
\ri)\ri] dx \nn \\
&=&
\frac{\hbar^2}{2m}\le[
\le(
\phi^\star(L) \psi'(L)- \psi(L) \phi'^\star(L)
\ri) -
\le(\phi^\star(0) \psi'(0) - \psi(0) \phi'^\star(0)
\ri)\ri] \nn \\
&& +
\frac{i\alpha \hbar^3}{m} \le[
\le(
\phi''^\star(L) \psi(L) + \phi^\star(L) \psi''(L) - \phi'^\star(L) \psi'(L)
\ri) \right. \nn \\
&&
\left. -\le(
\phi''^\star(0) \psi(0) + \phi^\star(0) \psi''(0) - \phi'^\star(0) \psi'(0)\ri)\ri]
\eea
\par\noindent
which for $\phi=\psi$ reduces to
\bea
 (H^\dagger \phi,\phi) - (\phi, H^\dagger \phi) &=& \frac{2i\hbar^2}{2m}\frac{1}{2i} 
\le[L
\le(
\phi^\star(L) \phi'(L) - \phi(L) \phi'^\star(L) \ri)
 - L \le( \phi^\star(0) \phi'(0) - \phi(0) \phi'^\star
\ri)
\ri] \nn \\
&& + \frac{i\alpha\hbar^3}{mL^2}
\le[
L^2 \le(
\phi''^\star \phi(L) + \phi^\star(L) \phi''(L)- \phi'^\star (L)\phi'(L)
\ri)
-L^2 \le(
\phi''^\star(0) \phi(0) + \phi^\star(0) \phi''(0) - \phi'^\star(0) \phi'(0)
\ri)
\ri] \nn \\
&=& \kappa \le[
\left(
|L \phi'(0) - i \phi(0) |^2 + |L \phi'(L) + i\phi(L) |^2 \right. .- |L\phi'(0)+i\phi(0)|^2 - |L\phi'(L)-i\phi(L)|  \right)
\nn \\
&&
+ \frac{\alpha\hbar}{2L} \le(
|L^2 \phi''(0) + \phi(0) |^2 + |L^2 \phi''(L) + \phi(L) |^2 \right.  - |L^2 \phi''(0) - \phi(0) |^2 - |L^2\phi''(L) - \phi(L)|^2 \nn \\
&& \left.  \left.
+ L^2 | \phi'(L)|^2 - L^2 |\phi'(0)|^2
\ri)
\ri]
\la{adj1}
\eea

\par\noindent
where $\kappa=i\hbar^2/mL$,
as in \cite{Bonneau:1999zq}. In addition, we have introduced factors of $L$ for dimensional reasons and
the following identities were employed
\bea
\frac{1}{2i}\le( x y^\star - y x^\star \ri) &=& \frac{1}{4} \le( |x+iy|^2 - |x-iy|^2 \ri) \nn \\
2\le( xy^\star+ y x^\star \ri) &=& |x+y|^2 - |x-y|^2 ~.\nn
\eea
At this point, a  couple of comments are in order. Firstly, it is evident that if $\alpha=0$, then Eq.(\ref{adj1}) reduces to Eq.(30) of \cite{Bonneau:1999zq}, as expected. Secondly, if we try to extend our computation by letting $L \rightarrow \infty$, 
then it is easily seen that Eq. (\ref{adj1}) blows up. This is a result that we were expecting since as we mentioned before (see 
section IV. A) when the domain of the Hamiltonian is  $\mathcal{D}(\hat{H}) = \mathcal{L}^{2}(0,\infty)$, the Hamiltonian operator is not a self-adjoint operator.\\

\par\noindent
To bring out the $U(3)$ invariance of the self-adjoint extensions of the Hamiltonian, defined by 
($H^\dagger \phi,\psi) - (\phi, H^\dagger \psi)$, one might, for example, use a $5$-dimensional
representation of $U(3)$ [we scale both sides of Eq.(\ref{adj1}) by $\kappa$, and in the following ,
$A=\alpha\hbar/2L$]
\bea
\left(
    \begin{array}{c}
      L\phi'(0)-i\phi(0) \\
      L\phi'(L)+ i\phi(L) \\
      A(L^2 \phi''(0) + \phi(0)) \\
      A(L^2 \phi''(L)+ \phi(L))  \\
      A \phi'(L)
    \end{array}
  \right)
  =
  U \left(
    \begin{array}{c}
      L\phi'(0)+ i\phi(0) \\
      L\phi'(L) - i\phi(L) \\
      A( L^2 \phi''(0) - \phi(0) )  \\
      A( L^2 \phi''(L)+ \phi(L)) \\
      A \phi'(0)
    \end{array}
  \right)~.
\eea
The construction of $U$ could be involved. However, note that corresponding to $U=I$, the $5 \times 5$ identity matrix, 
$\phi(0)=\phi(L)=0$ (like we assumed in \cite{Ali:2009zq}), 
$\phi''(0),\phi''(L)$ arbitrary, and
$|\phi'(0)|=|\phi'(L)|$ is {\it a} valid solution set
of $(H^\dagger \phi,\phi) - (\phi, H^\dagger \phi)=0$.
From \cite{Ali:2009zq}, we get ($\ell \equiv 2\alpha\hbar$)
\bea
 \phi' &=& iA \le[k' e^{ik'x} -\frac{1}{\ell} e^{ix/\ell}\ri] 
- iB \le[ k'' e^{-ik''x} + \frac{1}{\ell}~e^{ix/\ell} \ri] \nn \\
&=& 
2iAk \le[ \cos kx + \frac{ik\ell}{2}\sin kx \ri] - Ak^3 \ell x \cos kx \nn \\
&& + iC \le[ k e^{ikx} + \frac{1}{\ell}~e^{ix/\ell} \ri]~. 
\eea
using $A+B+C=0$, $k'=k(1+k\alpha\hbar)$ and $k''=k(1-k\alpha\hbar)$, 
and simplifying. 

\par\noindent
Then the condition $|\phi'(0)|=|\phi'(L)|$ translates to, 
\be
(-1)^n \le[
2iAk \cdot e^{i k^2 \ell L/2} + iC \le(k + \frac{1}{\ell}e^{i2t\pi} \ri)
\ri]
 = e^{i\chi} \le[
2iAk + iC \le( k + \frac{1}{\ell} \ri)
\ri] 
\ee

\par\noindent
where $2t\pi \equiv \pi(p-n) + 2 \epsilon_i \theta_C$, where $\epsilon_1=1$ and $\epsilon_2=0$,
corresponding to solutions (20) and (21) of \cite{Ali:2009zq} respectively, and $\chi$ is an arbitrary phase. 
This can be simplified to 
\bea
(-1)^n e^{i\phi} 
\le[ 
2iAk + \frac{iC}{\ell} e^{i(2t\pi -\Phi)}
\ri] 
=
e^{i\chi} \le[ 
2iAk + \frac{iC}{\ell}
\ri] \la{sol1}
\eea
where 
$\Phi = k^2\ell L/2$ and we have used $k + 1/\ell \approx 1/\ell$. 
The above admits of the solution 
\bea
\chi = n\pi + \Phi~,~~& 2t\pi -\phi = 2p_1 \pi~,~p_1 \in \mathbb{N}.
\eea
At this point, it is noteworthy that the above solution shows that the quantization conditions 
(20) and (21) of \cite{Ali:2009zq} form a subset of the general solution (likely with more parameters). 
\section{Conclusions}
\par\noindent
In this work we explore the self-adjointness of the GUP-modified momentum
and Hamiltonian operators over different domains. The reason for this investigation is that it is well-known that in Physics we are interested in observables with real eigenvalues, which are guaranteed only for self-adjoint operators, barring potentially pathological cases that are of no apparent interest to our purposes. Thus, it is very important to know if an operator is self-adjoint or not, or if it has self-adjoint extensions over specific domains. Outside these domains, eigenvalues are not real-valued, and hence the operators cannot be considered as physical operators. 
The domains under study are: (a) the whole real axis, (b) the positive semi-axis, 
and (c) a finite  internal. In order to utilize the von Neumann's theorem, we first obtain the functions $\psi_{\pm}$, 
second we compute the  dimensions $n_{\pm}$ of the deficiency subspaces of the GUP-modified momentum and 
Hamiltonian operators, and finally we infer whether the operators are self-adjoint or not, or they have infinitely many self-adjoint 
extensions. This analysis is adopted for all three cases of GUP-modified momentum and Hamiltonian operators, namely with a linear term 
in the GUP parameter $\alpha$, with a quadratic term in $\alpha$, and with both terms, i.e., the linear and the quadratic 
in $\alpha$, to be included. It is noteworthy that the GUP-modified momentum operator with both terms in $\alpha$ to be included is 
self-adjoint operator when its domain is the whole real axis, it is not self-adjoint operator when its domain is the positive semi-axis, and it has  
infinitely many self-adjoint extensions when its domain is a finite internal. Furthermore, the GUP-modified Hamiltonian operator with 
both terms in $\alpha$ to be included is self-adjoint operator when its domain is the whole real axis, and it has infinitely many
 self-adjoint extensions when its domain is the positive semi-axis or a finite internal. 
At this point, it should be stressed that the self-adjoint extensions of different domains are parametrized by different unitary 
groups. Finally, a simple example of the Hamiltonian for a particle in a box is given and the solutions for the boundary 
conditions which describe the self-adjoint extensions of the specific operator are obtained. 
\section*{Acknowledgments}
\par\noindent
We thank Jorma Louko for fruitful comments and  Galliano Valent for useful correspondence. In addition, we thank the anonymous referee for his helpful comments. This work is supported by the Natural Sciences and Engineering Research Council of Canada.





\end{document}